\documentclass[letterpaper, english, aps, prl, reprint, twocolumn, superscriptaddress]{revtex4-2}
\usepackage[T1]{fontenc}
\usepackage{color}
\usepackage{amsmath}
\usepackage{amssymb}
\usepackage{graphicx}
\usepackage{multirow}
\usepackage{booktabs}
\makeatletter

\usepackage[bookmarks=true,colorlinks,linkcolor=black,anchorcolor=green,citecolor=blue,urlcolor=blue]{hyperref}

\renewcommand{\fnum@figure}{FIG. \thefigure}
\renewcommand{\fnum@table}{TABLE. \thetable}

\makeatother

\usepackage{babel}
\begin{document}

\title{$C_n$-symmetric higher-order topological crystalline insulators in atomically thin transition-metal dichalcogenides}

\author{Shifeng Qian}
\affiliation{Centre for Quantum Physics, Key Laboratory of Advanced Optoelectronic Quantum Architecture and Measurement (MOE), School of Physics, Beijing Institute of Technology, Beijing, 100081, China}
\affiliation{Beijing Key Lab of Nanophotonics \& Ultrafine Optoelectronic Systems, School of Physics, Beijing Institute of Technology, Beijing, 100081, China}

\author{Gui-Bin Liu}
\affiliation{Centre for Quantum Physics, Key Laboratory of Advanced Optoelectronic Quantum Architecture and Measurement (MOE), School of Physics, Beijing Institute of Technology, Beijing, 100081, China}
\affiliation{Beijing Key Lab of Nanophotonics \& Ultrafine Optoelectronic Systems, School of Physics, Beijing Institute of Technology, Beijing, 100081, China}

\author{Cheng-Cheng Liu}
\email{ccliu@bit.edu.cn}
\affiliation{Centre for Quantum Physics, Key Laboratory of Advanced Optoelectronic Quantum Architecture and Measurement (MOE), School of Physics, Beijing Institute of Technology, Beijing, 100081, China}
\affiliation{Beijing Key Lab of Nanophotonics \& Ultrafine Optoelectronic Systems, School of Physics, Beijing Institute of Technology, Beijing, 100081, China}

\author{Yugui Yao}
\affiliation{Centre for Quantum Physics, Key Laboratory of Advanced Optoelectronic Quantum Architecture and Measurement (MOE), School of Physics, Beijing Institute of Technology, Beijing, 100081, China}
\affiliation{Beijing Key Lab of Nanophotonics \& Ultrafine Optoelectronic Systems, School of Physics, Beijing Institute of Technology, Beijing, 100081, China}

\begin{abstract}
Based on first-principles calculations and symmetry analysis, we predict atomically thin ($1-N$ layers) 2H group-VIB TMDs $MX_2$ ($M$ = Mo, W; $X$ = S, Se, Te) are large-gap higher-order topological crystalline insulators protected by $C_3$ rotation symmetry. We explicitly demonstrate the nontrivial topological indices and existence of the hallmark corner states with quantized fractional charge for these familiar TMDs with large bulk optical band gaps ($1.64-1.95$ eV for the monolayers), which would facilitate the experimental detection by STM. We find that the well-defined corner states exist in the triangular finite-size flakes with armchair edges of the atomically thin ($1-N$ layers) 2H group-VIB TMDs, and the corresponding quantized fractional charge is the number of layers $N$ divided by 3 modulo integers, which will simply double including spin degree of freedom. 
\end{abstract}
\maketitle

\paragraph{\textcolor{blue}{Introduction.}\textemdash{}}

Atomically thin two-dimensional (2D) transition-metal dichalcogenide (TMD) semiconductors have attracted great scientific and technological interest because of the extraordinary properties, such as the direct band gap in the visible frequency range, remarkable optical properties, and rich valley-related physics \cite{Heinz2010, Wang2010, Xiao2011, Feng2012, Liu2015,Wu2019,BilayersTMD}. Among various TMDs, the group-VIB ones $MX_2$ ($M$ = Mo, W; $X$ = S, Se, Te) have been most extensively studied in 2D forms, where both the monolayers and few-layers are proved to be stable in air at room temperature except $M$Te$_2$.  

Exploring new topological quantum states, especially high-order topological crystal insulator (HOTCI) states, and discovering good candidate materials are among the most active studies in condensed matter physics and materials science \cite{TIRev1, TIRev2, po2017symmetry,ebrs, Slager2017PRX, fenlei1, fenlei2, fenlei3, HOTI2, HOTI3, HOTI4, HOTI1, HOTI5,  HOTIREV,  BeathKogome, HOTIA, HOTIGeneral, HOTIPC, Fengliu, Fengliugraphene}. Different from a conventional TCI having protected gapless states on its symmetric boundaries with one dimension lower than the bulk, HOTCIs feature the lower-dimensional protected boundary states. For instance, three dimensional (3D) second-order TCIs host 1D gapless states along their hinges, and 2D HOTCIs display in-gap corner modes.  A few theoretical materials for 3D and 2D HOTCIs are proposed in the literature \cite{V_Topo, graphdiyne1, graphdiyne2, graphyne1,graphyne2, TBG1, TBG2, MnBiTe,HOTI, MnBiTe, XiDai1, XiDai2, Chen2020, Our_arXiv2021, Huang_arXiv2021}. However second-order TCIs have only been experimentally observed in 3D single crystal bismuth \cite{HOTI_Bi} and some artificial systems \cite{mechanic, acoustic1, acoustic2, photonic1, photonic2, photonic3, elect1, elect2, elect3, micro, HOTISonic1, HOTIElast}. Therefore, proposing and discovering ideal and real material candidates of 2D HOTCIs are still urgent and important. 

\begin{figure}
	\begin{center}
		\includegraphics[width=1\linewidth]{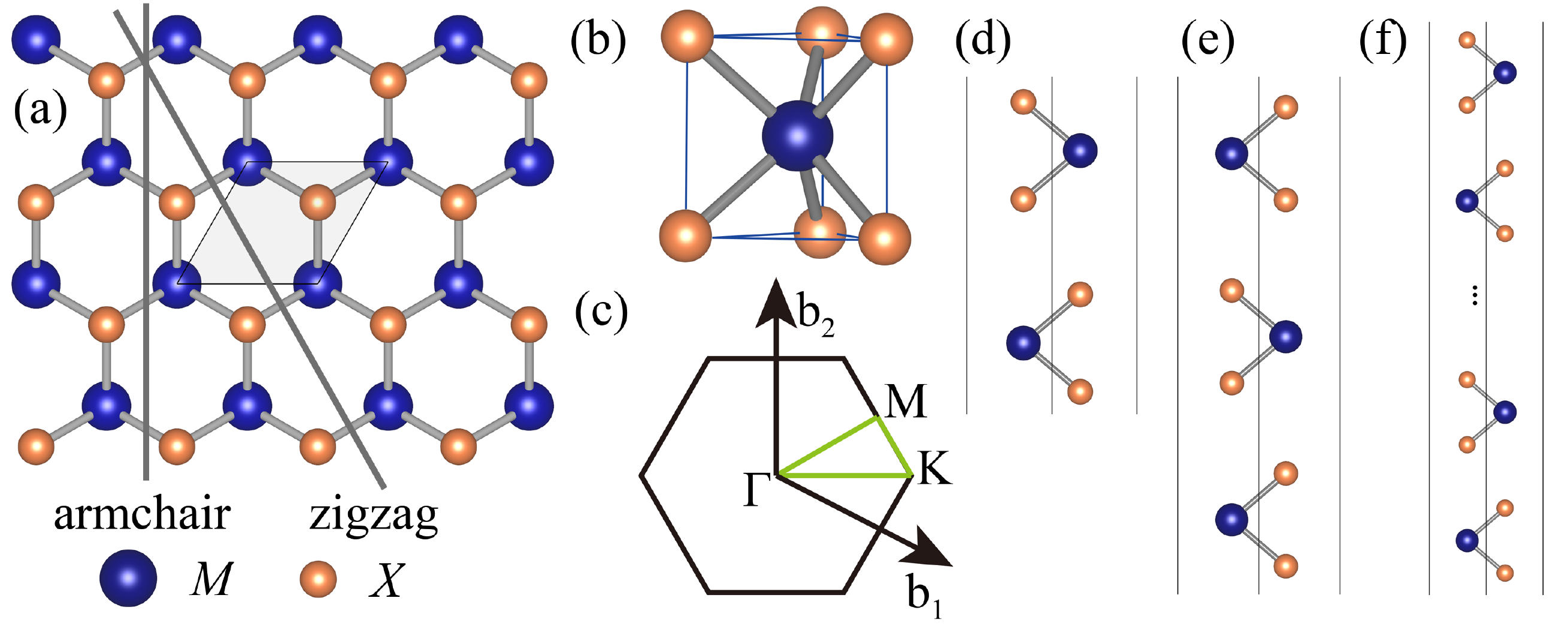}
	\end{center} 
	\caption{(a) Top view of transition metal dichalcogenides $MX_2$ ($M$ = Mo, W; $X$ = S, Se, Te) monolayer. The blue and orange spheres represent $M$ and $X$ atoms respectively. The light gray diamond region is the unit cell of $MX_2$. The gray lines mark the zigzag and armchair edges. (b) Trigonal prismatic coordination geometry of MX$_2$. (c) The first Brillouin zone with high-symmetric points. $\mathbf{b_1}$ and $\mathbf{b_2}$ are the reciprocal lattice vectors. (d) (e) (f) Side view of bilayer, trilayer and multilayer $MX_2$ of 2H stacking.}  \label{fig:1}
\end{figure}

We have put forward guidelines for designing the 2D HOTCI state in hexagonal lattices with $s$ and $p$ orbitals, and predicted abundant material candidates \cite{Our_arXiv2021}, recently. In this work we generalize to the systems of $d$ orbitals, among which the atomically thin group-VIB TMDs (monolayers, bilayers, few-layers, etc.) are most representative. We take 2H bilayer group-VIB TMDs $MX_2$ ($M$ = Mo, W; $X$ = S, Se, Te) with inversion symmetry (\textit{P}) and time-reversal symmetry (\textit{T}) as the starting point. The band topology is characterized by the second Stiefel-Whitney class with \textit{P} and \textit{T} symmetry. Although the second Stiefel-Whitney number $w_{2}$ is zero for the bilayers, there are still nonzero topological index and corner states with quantized fractional charge protected by $C_3$ rotation symmetry localized at the corners of the 2H bilayer group-VIB TMDs, indicating a HOTCI, by combined density function theory (DFT) simulation and symmetry analysis. Actually, the rotation symmetry protected HOTCI states do not need the inversion symmetry but $C_3$, which is owned by all 2H odd-layer group-VIB TMDs (monolayers and trilayers, etc) and other 2H even-layer group-VIB TMDs (quadlayers, etc). We find other 2H group-VIB TMDs (monolayers, trilayers, quadlayers, etc.) are also large-gap HOTCIs protected by $C_3$ symmetry. These atomically thin group-VIB TMDs $MX_2$ ($M$ = Mo, W; $X$ = S, Se, Te) have large bulk optical band gap (e.g., $1.64-1.95$ eV for the monolayers), which would facilitate the experimental verification and exploration of the HOTCI phases. 

\paragraph{\textcolor{blue}{HOTCIs in 2H bilayer group-VIB TMDs with \textit{PT} and $C_3$ symmetries.}\textemdash{}} The bulk of group-VIB TMDs with 2H stacking crystalizes in space group $D_{6h}^4$ with inversion symmetry. For the atomically thin counterparts, the symmetry is reduced to $D_{3h}$ (monolayers), $D_{3d}$ (bilayers), $D_{3h}$ (trilayers), and $D_{3d}$ (quadlayers), as shown in Fig. \ref{fig:1}. The inversion symmetry is preserved in the even-layer films ($D_{3d}$), but broken in the odd-layer ones ($D_{3h}$). First, we take into account the centrosymmetric spinless bilayer, with spin-orbital coupling (SOC) discussed later. The higher-order band topology of spinless systems with $\textit{P}$$\textit{T}$ symmetry can be characterized by the so-called second Stiefel-Whitney number $w_{2}$ (See details in Supplemental Material  \cite{SuppMater}). We take bilayer MoS$_{2}$ as example, and calculate the second Stiefel-Whitney number $w_{2}$, and find $w_{2}=0$ [Fig. \ref{fig:2} (b)], which is also checked by using parity criterion and nested Wilson loop (See details in Supplemental Material \cite{SuppMater}), seeming to indicate a trivial state. However, we still find the nontrivial topological indices and the hallmark corner states with fractional charge protected only by $C_3$ symmetry in the bilayer. 

\begin{figure}
	\begin{center}
		\includegraphics[width=1\linewidth]{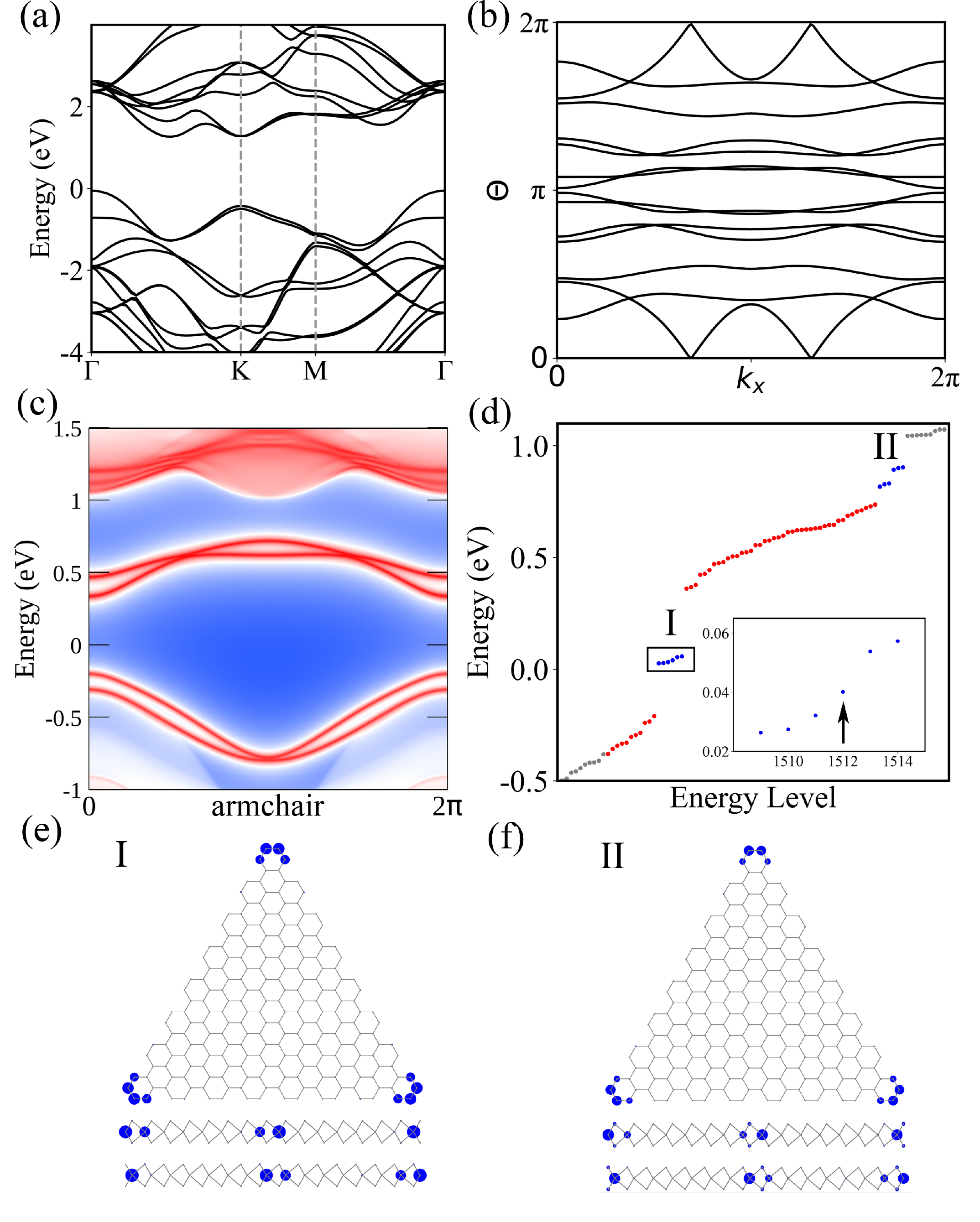}
	\end{center}
	\caption{(a) (c) Bulk band and edge states of bilayer MoS$_2$. (b) Wilson loop of bilayer MoS$_2$. The number of crossings on $\theta = \pi$ is zero, indicating the $w_2=0$. (d) Energy spectrum of a triangular finite-size flake of bilayer MoS$_2$ calculated from DFT and Wannier function. The gray, red and blue dots mark the respective bulk, edge, and corner states. The corner states are split in two groups labeled by I and II in different energy regions. The arrow denotes the Fermi level by valence electron counting. (e) (f) The top view and side view of the real-space distribution of the two groups of the corner states.} \label{fig:2}
\end{figure}

The nontrivial topological indices and hallmark corner states with quantized fractional charge can be described by the $C_3$-symmetry eigenvalues of the occupied energy bands at the high-symmetric points of the Brillouin zone (BZ) \cite{kekulemodel3}. The three-fold rotation symmetry eigenvalues at the high-symmetric points $\mathbf{\Sigma}^{(3)}$ in the BZ is denoted as $\Sigma_{m}^{(3)}=e^{2 \pi i(m-1) / 3}, \quad$ for $m=1,2, 3$. The topological invariants can be defined through the rotation eigenvalues at $\mathbf{\Sigma}^{(3)}$ compared to a certain reference point $\boldsymbol{\Gamma}=(0,0)$, i.e., $\left[\Sigma_{m}^{(3)}\right] \equiv \# \Sigma_{m}^{(3)}-\# \Gamma_{m}^{(3)}$, where $\# \Sigma_{m}^{(3)}$ ($\# \Gamma_{m}^{(3)}$) is the number of occupied bands with eigenvalue $\Sigma_{m}^{(3)}$ ($\Gamma_{m}^{(3)}$). The topological indices $\chi^{(3)}$ and fractional corner charge $Q_{\text {corner }}^{(3)}$ for $C_3$-symmetric HOTCIs read
\begin{equation}\label{Frac_Q}
	\begin{aligned}
\chi^{(3)}&=\left(\left[K_{1}^{(3)}\right], \left[K_{2}^{(3)}\right]\right),  \\
Q_{\text {corner}}^{(3)}&=\frac{e}{3}\left[K_{2}^{(3)}\right] \bmod e, 
	\end{aligned}
\end{equation}
where the superscript $3$ of $\chi^{(3)}$ and $Q_{\text {corner }}^{(3)}$ labels the $C_3$ symmetry, and $e$ is the charge of a free electron.

 \begin{figure*}
	\begin{center}
		\includegraphics[width=1\linewidth]{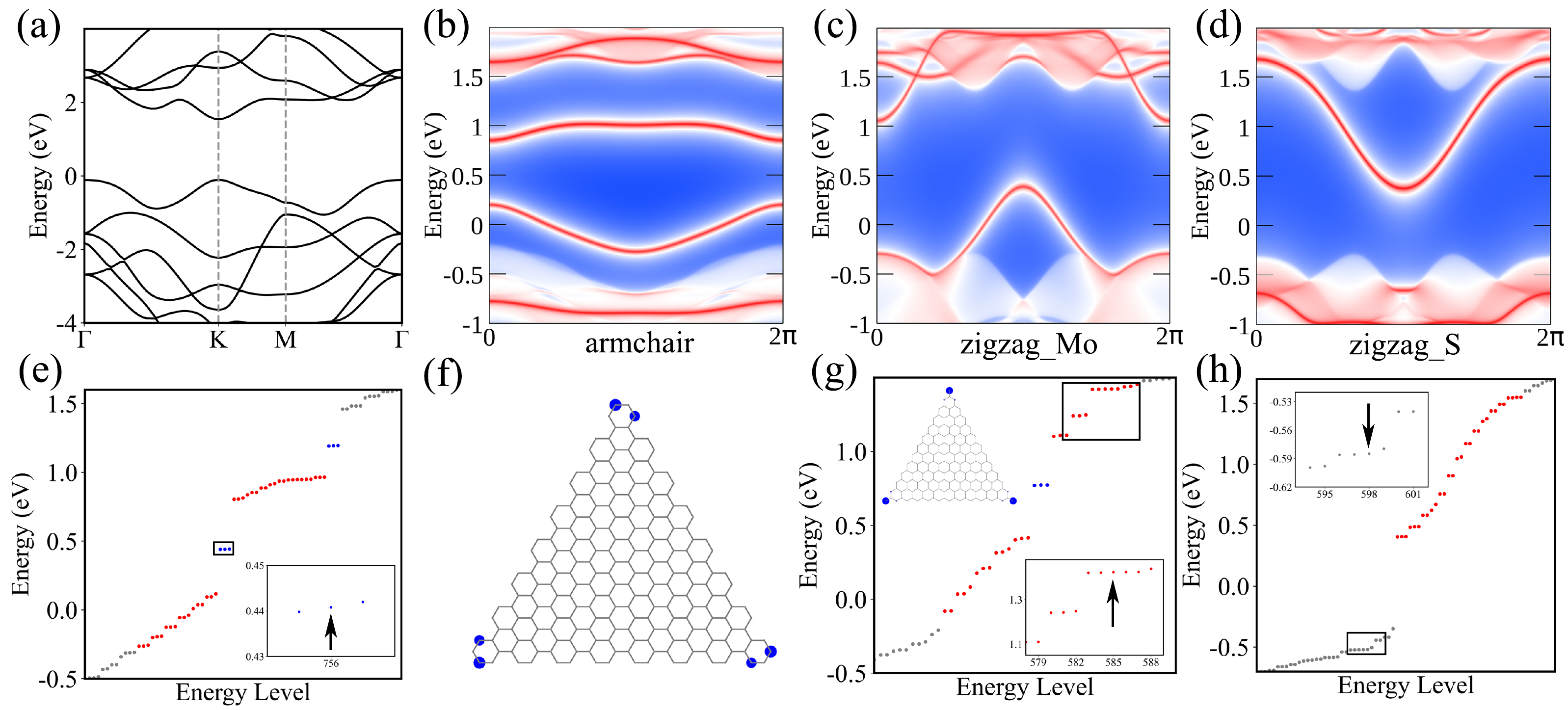}
	\end{center}
	\caption{(a) Bulk band of monolayer MoS$_2$. (b)-(d) The edge states of three semi-infinite planes with respective armchair, zigzag\_Mo, and zigzag\_S edges, calculated by DFT and Wanner function. (e) The energy spectra of a triangular nanoflake (f) with armchair edges calculated by DFT and Wanner function. The gray, red and blue dots in (e) stand for the respective bulk, edge, and corner states. The real-space distribution of the corner states framed by a rectangle is plotted in (f). The arrow in the zoomed rectangle marks the position of Fermi level by electron counting. (g) The energy spectra of a triangular nanoflake with zigzag\_Mo edges. Although there are still corner states in blue located at the corner, the Fermi level marked by an arrow and determined by electron counting is in the edge states in red, indicating an intrinsic metallic edge state. (h) The energy spectra of a triangular nanoflake with zigzag\_S edges. There are no in-gap corner states, and the Fermi level labeled by an arrow is located in the bulk states, signing an intrinsic metallic bulk state.} \label{fig:3}
\end{figure*}

We take the bilayer MoS$_{2}$ as an example, whose nonzero topological index is $\chi^{(3)}=\left(-6, 2\right)$ and the fractional corner charge $Q_{\text {corner}}^{(3)}=2e/3$. The bulk band structure and armchair edge states of bilayer MoS$_2$ are plotted in Figs. \ref{fig:2} (a) and (b). To explore the hallmark corner states of bilayer MoS$_{2}$, we calculate the energy spectrum of a triangular finite-size flake [Fig. \ref{fig:2}(d)]. There are two groups of states in blue, with one group near zero energy in the spectrum and the other with higher energy, labeled as I and II. We consider (Mo:$5s^14d^5$) and (S:$3p^4$) as the valence electron configuration of MoS$_{2}$. In the bilayer, one unit cell has two Mo atoms and four S atoms, so the number of the valence electrons is $2\times(6+2\times4)/2=14$, where the first multiplier 2 is for the layer number, the second multiplier 2 stands for two S atoms in the formula MoS$_{2}$, and the divisor 2 denotes the spinless case. The triangular flake consists of 108 unit cells, and has $108\times14=1512$ valence electrons. The Fermi level is determined by the electron counting and indicated by an arrow [Fig. \ref{fig:2}(d)]. The other group-VIB TMDs have the similar valence electron configuration and same valence electron counting. For the group I, the top view and side view of the real-space distribution of the six states are shown in Fig. \ref{fig:2}(e). From the top view, one can see these states are well located at the corners of the flake, corresponding to the corner states. From the side view, these corner states are evenly distributed on two layers, and mainly on the Mo atoms. Figure \ref{fig:2}(f) shows the top view and side view of the real-space distribution of the group II, with similar distribution as that of the group I. The similar analysis and results for the other bilayer group-VIB TMDs are given in Supplemental Material \cite{SuppMater}.

\paragraph{\textcolor{blue}{HOTCIs in monolayer group-VIB TMDs protected by $C_3$ symmetry.}\textemdash{}} The above $C_n$ symmetry protected HOTCI mechanism can directly apply to the monolayer ones. For the monolayer MoS$_{2}$, the nonzero topological index is $\chi^{(3)}=\left(-3, 1\right)$ and the fractional corner charge $Q_{\text {corner}}^{(3)}=e/3$. Figure \ref{fig:3} (a) shows the bulk band structure of monolayer MoS$_2$.  As shown in Fig. \ref{fig:1}(a), one usually cuts the group-VIB TMDs $MX_2$ with armchair and zigzag edges, and the zigzag edge has two versions differentiated by $M$ or $X$ termination, labeled by zigzag$\_M$ and zigzag$\_X$. As shown in Fig. \ref{fig:3}, we construct three semi-infinite planes and three triangular finite-size flakes with armchair, zigzag$\_M$ and zigzag$\_X$ edges, respectively. There are two gapped armchair edge states in the bulk gap, as shown in Fig. \ref{fig:3} (b). The edge states for zigzag\_Mo and zigzag\_S are plotted in Figs. \ref{fig:3} (c) and (d). To explicitly reveal the hallmark corner states, we calculate the energy spectrum of the three triangular nanoflakes with three different edges based on DFT calculation and Wannier function, as shown in Figs. \ref{fig:3} (e) (f), Fig. \ref{fig:3} (g), and Fig. \ref{fig:3} (h). For the armchair nanoflake, in the middle of the edge states exit three degenerate states (blue dots in Fig. \ref{fig:3}(e)), whose charge real-space distribution is plotted in Fig. \ref{fig:3}(f). Such three states are well located at the three corners of the flake [Fig. \ref{fig:3} (f)], i.e., the corner states. By similar electron counting as that of the bilayer, we fix the position of Fermi level, just right at the blue corner states, marked by an arrow in the zoomed rectangle in Fig. \ref{fig:3}(e). As for the two zigzag nanoflakes, the Fermi level is in the edge or bulk states with the hallmark corner states covered, as shown in Fig. \ref{fig:3} (g) and (h). We would like to stress that only the armchair flakes have appropriate electron filling, i.e., with corner states at the Fermi level, while the nanoflakes with two kinds of zigzag edges have metallic bulk or edge states. Therefore, we are mainly interested in the 1D or 0D samples with armchair edges for the group-VIB TMDs $MX_2$. The other monolayer group-VIB TMDs are also HOTCIs (See details in Supplemental Material \cite{SuppMater}).

\begin{figure}
	\begin{center}
		\includegraphics[width=1\linewidth]{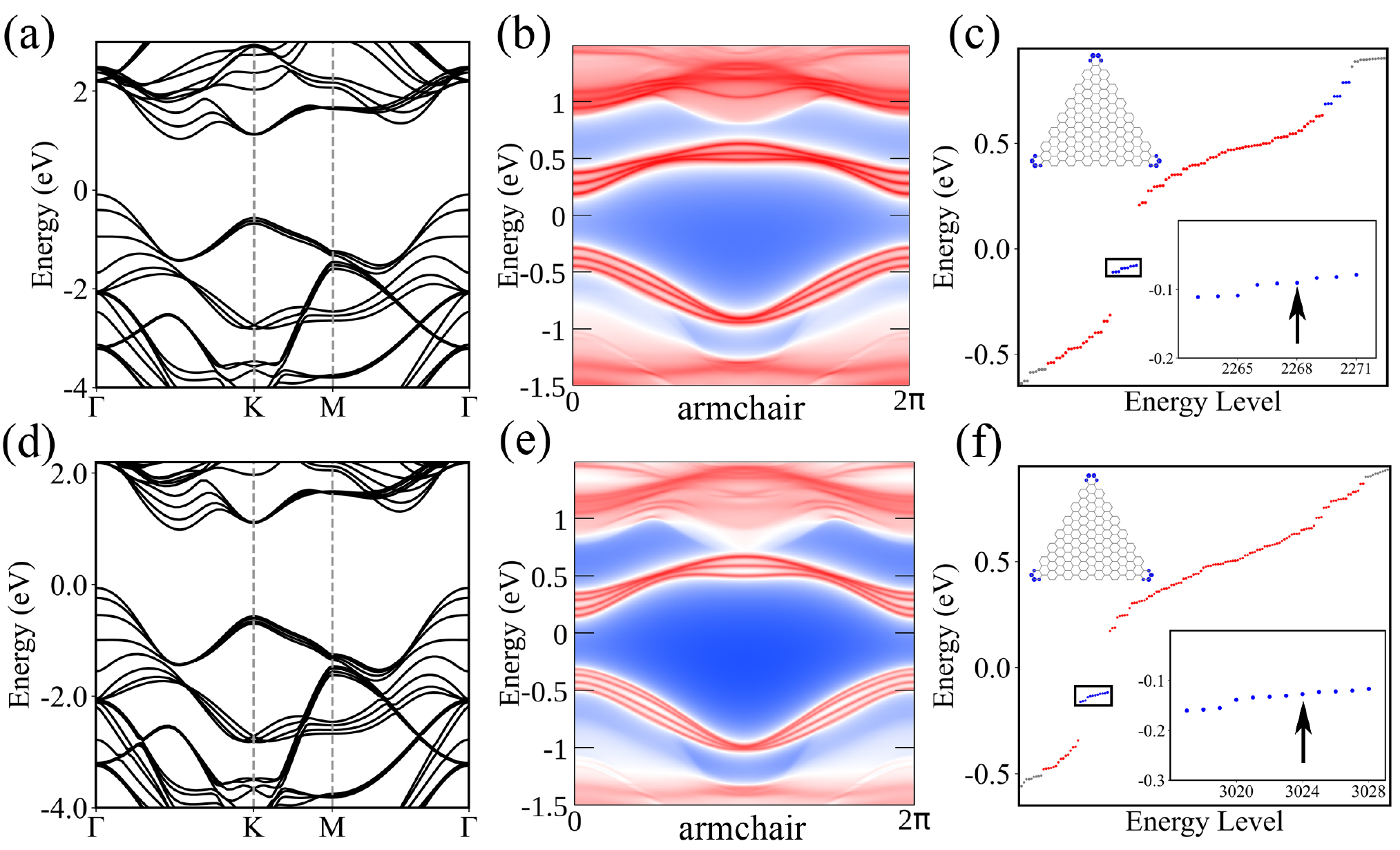}
	\end{center}
	\caption{(a) (d) Bulk band structures of trilayer and quadlayer MoS$_2$. (b) (e) Armchair edge states of trilayer and quadlayer MoS$_2$. Energy discrete spectra of the finite-size triangular flakes of trilayer (c) and quadlayer (f) MoS$_2$ calculated from DFT and Wannier function. The blue states in (c) and (f), whose charge real-space distribution is plotted in the insets, are well localized at the corners. The gray, red and blue dots represent the respective bulk, edge, and corner states. The Fermi level marked by the arrows in (c) and (f) is determined by electron counting.} \label{fig:4}
\end{figure}

\paragraph{\textcolor{blue}{HOTCIs in other few-layer group-VIB TMDs protected by $C_3$ symmetry.}\textemdash{}} Figures \ref{fig:4} (a) (d) show the bulk band structures of trilayer and quadlayer MoS$_2$. Their armchair edge states are plotted in Figs. \ref{fig:4} (b) (e). The discrete energy spectra of two triangular finite-size flakes of the trilayer and quadlayer are shown in Figs. \ref{fig:4} (c) (f). One can see the well-defined corner states are located at the corners of the both flakes. According to Eq. (\ref{Frac_Q}), the fractional corner charge of the quadlayer nanoflake is $Q_{\text {corner}}^{(3)}=e/3$. In spite of the clear corner states for the trilayer nanoflake, the corner charge is zero. In the trilayer (or 6, 9, etc., layers) nanoflake, the original corner states with $e/3$ fractional corner charge in each layer will mix together with an integer corner charge left, thus without fractional corner charge. Based on the above analysis, we propose a simple formula to calculate the fractional corner charge of the few-layer TMDs, which reads
\begin{equation}
Q_{N}^{\text {corner }}=e\frac{N}{3} \bmod e, 
\end{equation}
where $N$ is the number of layers. The similar analysis and results hold for the other trilayer and quadlayer group-VIB TMDs as well as the other few-layer group-VIB TMDs.

\begin{figure}
	\begin{center}
		\includegraphics[width=1\linewidth]{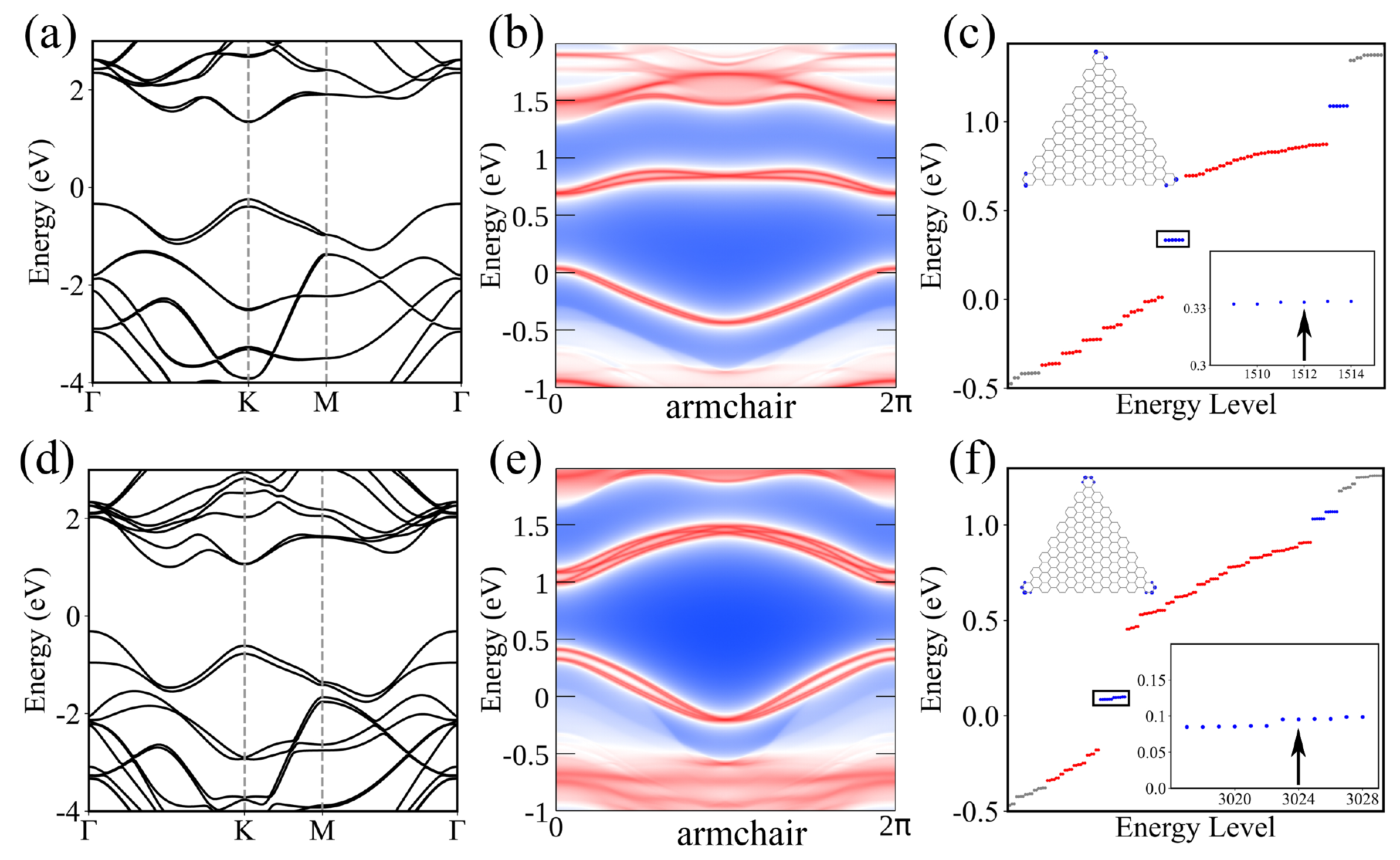}
	\end{center}
	\caption{(a) (d) Bulk band structures with spin-orbital coupling (SOC) of monolayer and bilayer MoS$_2$. (b) (e) Armchair edge states with SOC of monolayer and bilayer MoS$_2$. Energy discrete spectra with SOC of the finite-size triangular flakes of monolayer (c) and bilayer (f) MoS$_2$ calculated from DFT and Wannier function. The corresponding charge spatial distribution of blue states in (c) and (f) plotted in their insets are well localized at the corners. The gray, red and blue dots stand for the respective bulk, edge, and corner states. The Fermi level marked by the arrows in (c) and (f) is determined by electron counting.} \label{fig:5}
\end{figure}

\paragraph{\textcolor{blue}{The effect of spin-orbital coupling.}\textemdash{}} The fractional corner charge in 2D system with SOC can also be classified by rotation symmetry \cite{V_Topo}. The topological indices $\chi^{(3)}$ and fractional corner charge $Q_{\text {corner }}^{(3)}$ for $C_3$-symmetry protected HOTCIs with SOC read
\begin{equation}
	\begin{aligned}
		\chi^{(3)}&=\left(\left[K_{1}^{(3)}\right], \left[K_{2}^{(3)}\right]\right),  \\
		Q_{\text {corner }}^{(3)}&=\frac{2e}{3}(\left[K_{1}^{(3)}\right]+\left[K_{2}^{(3)}\right]) \bmod 2e.
	\end{aligned}
\end{equation}
We take the monolayer and bilayer MoS$_2$ as examples, whose nonzero topological index $\chi^{(3)}$ are (-2, 3) and (-4, 6) respectively and the fractional corner charge  $Q_{\text {corner}}^{(3)}$ are $2e/3$ and $4e/3$ respectively. The band structures and edge states of monolayer and bilayer MoS$_2$ (Fig. \ref{fig:5}) undergo some changes by the SOC effect, but the topology of these materials do not changed. We calculate the discrete energy spectrum of triangular finite-size flakes, as shown in of Figs. \ref{fig:5}(c) and (f).  There are also existing two groups of corner states near zero energy in the spectrum and the other with higher energy, which similar to the spinless condition. However, the number of corner states of each group doubled compared to the spinless case. The corresponding charge distribution of corner states are plotted in their inset. Therefore, the HOTCI topology of $MX_2$ is also preserved in the SOC case since the large bulk band gap dominates the SOC effect. Compared with the spinless case, the corner charge will double when SOC taken into account.

\paragraph{\textcolor{blue}{Conclusion and discussion.}\textemdash{}} We have demonstrated that $C_n$-symmetric large-gap higher-order topological crystalline insulators in atomically thin group-VIB TMDs $MX_2$ ($M$ = Mo, W; $X$ = S, Se, Te), whose monolayer and few-layer samples can be prepared from the bulk counterparts by using a mechanical exfoliation technique similar to that employed for graphene \cite{Heinz2010}. The nontrivial higher-order topology of these TMDs is revealed by the nonzero topological indices and existence of the hallmark corner states with quantized fractional charge. The $C_n$-symmetric HOTCIs in atomically thin group-VIB TMDs with large optical gaps (about 1.8 eV) would facilitate the experimental detection of the hallmark corner states as sharp peaks in the scanning tunneling spectroscopy (STS) measurement, when the scanning tip approaches the corners. As the atomically thin group-VIB TMDs are easy to produce with high quality, they are ideal material candidates to explore the HOTCI states and the related remarkable properties. 

When the number of layers $N$ of the group-VIB TMDs is large, the many-layer systems will approach the 3D bulks, and we find the corner states could evolves into hinge states \cite{Ourwork_In_Preparation}. In addition, the planes of $k_z=0$ and $k_z=\pi$ have the same nonzero topological indices protected by $C_3$ symmetry in the 3D group-VIB TMDs, which suggests that the 3D 2H group-VIB TMDs are $C_n$-symmetric higher-order weak topological crystalline insulators \cite{Ourwork_In_Preparation}. These contents are beyond the scope of this work, and left for our next work.

\begin{acknowledgments}
S. Qian and C.-C. Liu are supported by National Key R\&D Program of China (Grant No. 2020YFA0308800), and NSF of China (Grants No. 11922401, No. 11774028). Y.  Yao is supported by National Key R\&D Program of China (Grant No. 2020YFA0308800), NSF of China (Grants No. 12061131002, No. 11734003), and Strategic Priority Research Program of Chinese Academy of Sciences (Grant No. XDB30000000). G.-B. Liu is supported by National Key R\&D Program of China (Grant No. 2017YFB0701600) and Beijing Natural Science Foundation (Grant No. Z190006).
\end{acknowledgments}

\paragraph{Note added.} We become aware of an independent work on arXiv recently \cite{Xie_arXiv2021}. The work proposes second-order topological insulators in monolayer group-VIB TMDs, and the results of the monolayer group-VIB TMDs are consistent with ours.

\bibliographystyle{apsrev4-2}
\bibliography{reference}

\end{document}